\documentclass[11pt]{article}

\usepackage[a4paper,margin=1in]{geometry}
\usepackage{graphicx}
\usepackage[hidelinks]{hyperref}
\usepackage{amsmath}
\usepackage{amssymb}
\usepackage{bm}
\usepackage{longtable}
\usepackage{booktabs}
\usepackage{multirow}
\usepackage{ragged2e}
\usepackage{indentfirst}
\usepackage{makecell}

\begin{document}
\justifying

\title{High-Throughput Normalized Min-Sum Belief Propagation Decoding for Quantum LDPC Codes with Near-Memory Processing}

\author{Jeonggeun Seo, Youngsun Han, Leanghok Hour, and Dongmin Kim\thanks{Corresponding author: kdm902077@pukyong.ac.kr}\\
\small Pukyong National University, Busan, Republic of Korea}
\date{}

\maketitle

\begin{abstract}
\justifying

Real-time quantum error correction requires classical decoders to sustain increasing syndrome-processing demands as fault-tolerant quantum systems scale while maintaining low and predictable latency within each error correction cycle. For quantum low-density parity-check (qLDPC) codes decoded using iterative belief propagation (BP), repeated message updates over sparse Tanner graphs generate substantial memory-access and data movement demands, making data locality and parallel processing important for scalable decoding. Motivated by these workload characteristics, we map normalized Min-Sum BP decoding of the \([[144,12,12]]\) Bivariate Bicycle qLDPC code onto a DPU-based Processing-in-Memory (PIM) architecture that enables near-memory processing. The proposed design assigns 11 tasklets to cooperatively decode a single syndrome within each DPU, while multiple DPUs process independent syndrome instances in parallel. We evaluate the proposed decoder using uPIMulator under a data qubit Pauli error model with ideal syndrome measurements and compare its throughput, per-syndrome processing time, logical error rate (LER), and single syndrome tail latency with a 16-logical-CPU baseline. At a component-wise physical error probability of \(p=0.001\) and one BP iteration, the projected aggregate kernel throughput of a 2,560-DPU configuration reaches \(1.071\times10^{7}\) decodes/s, compared with \(1.22\times10^{6}\) decodes/s for the CPU baseline, corresponding to an \(8.8\times\) throughput improvement. From two BP iterations onward, the measured LER remains below the corresponding physical error probability across all evaluated values of \(p\). For one to five BP iterations, the maximum sampled serialized \(X+Z\) DPU compute latency remains below the \(1~\mathrm{ms}\) decoder-side timing reference used for trapped-ion QEC, reaching approximately \(0.873~\mathrm{ms}\) at five iterations. The reported latency represents decoder-kernel compute time, and the results show that the proposed architecture combines high projected aggregate throughput with sub-millisecond decoding latency under the evaluated conditions. These results demonstrate the potential of near-memory processing for high-throughput qLDPC BP decoding.

\end{abstract}


\vspace{1em}

\section{Introduction}
Fault-Tolerant Quantum Computing (FTQC) aims to execute quantum algorithms reliably in the presence of physical noise~\cite{terhal2015quantum,preskill1998reliable}. In practical quantum devices, quantum information is continuously affected by gate errors, measurement errors, and idle errors~\cite{preskill1998reliable,fowler2012surface,cross2009comparative}. Quantum Error Correction (QEC) suppresses the accumulation of these errors by encoding logical information into multiple physical qubits and repeatedly extracting error syndromes without directly measuring the encoded logical state~\cite{terhal2015quantum,fowler2012surface}. In each QEC round, a classical decoder processes the measured syndrome, estimates an error configuration, and determines the corresponding correction or Pauli-frame update~\cite{skoric2023parallel,das2022afs,battistel2023realtime}. As FTQC systems scale, decoders must continuously process increasing volumes of syndrome information within the available QEC-cycle timing budget, making decoding accuracy, throughput, and latency important requirements for scalable QEC~\cite{skoric2023parallel,das2022afs,ye2025beam}.

Quantum Low-Density Parity-Check (qLDPC) codes have attracted significant attention for scalable fault-tolerant quantum computation because their sparse parity-check structures can support higher encoding rates and potentially reduce physical qubit overhead~\cite{breuckmann2021quantum,tillich2014quantum,bravyi2024high}. Their sparse parity-check matrices can be represented as Tanner graphs composed of variable nodes and check nodes~\cite{kschischang2001factor,Tanner1981recursive}. This structure is naturally suited to iterative Belief Propagation (BP) decoding, where neighboring nodes exchange reliability messages to estimate an error configuration~\cite{kschischang2001factor,poulin2008iterative,yao2023belief}. During each BP iteration, edge-associated messages and node reliabilities are repeatedly read, computed, and updated, and this message state is retained and revisited across successive iterations~\cite{mackay1999good,richardson2001design}. This execution pattern distinguishes BP from decoding approaches centered on candidate search or post-processing, whose primary computational kernels follow different data access patterns~\cite{ye2025beam,valls2021syndrome}. Consequently, continuous BP decoding combines sparse arithmetic with repeated memory access and data movement, making memory locality an important architectural consideration. At the same time, local Tanner-graph updates and independent syndrome instances expose parallelism both within individual decoding tasks and across multiple tasks. These characteristics motivate near-memory processing, where frequently accessed decoding state can remain close to the processing units while large-scale parallelism is exploited through a Processing-in-Memory (PIM) architecture~\cite{mutlu2019processing,gomezluna2022benchmarking,baumstark2023accelerating,hyun2024pathfinding}.

Research on qLDPC decoding has progressed through both algorithmic development and hardware acceleration. BP-based decoders have been extended with methods such as guided decimation to improve decoding performance and convergence behavior~\cite{yao2023belief}. Hardware-oriented studies have investigated FPGA-, ASIC-, and GPU-based implementations of Min-Sum, BP, OSD-assisted, and related qLDPC decoding workloads~\cite{valls2021syndrome,bascones2025exploring,ferraz2025gpuqldpc}. In parallel, PIM architectures have been applied to classical LDPC decoding, including bit-flipping and non-binary LDPC decoders, demonstrating the feasibility of executing LDPC decoding operations close to memory~\cite{ferraz2024inmemorybf,ferraz2025inmemorynbldpc}. These developments motivate extending near-memory execution to the iterative message-passing workload of qLDPC BP decoding. In this work, we develop a DPU-based PIM implementation of normalized Min-Sum BP in which frequently accessed decoding state is maintained near the DPU processing pipeline. Within each DPU, 11 tasklets divide the check-node update workload of a single syndrome instance and synchronize at color-group boundaries, while independent syndrome instances are processed in parallel across multiple DPUs. This execution structure targets high aggregate decoding throughput together with predictable per-syndrome latency.

The main contributions of this work are summarized as follows:

\begin{itemize}
    \item We propose a DPU-based near-memory execution scheme for normalized Min-Sum BP decoding of the \([[144,12,12]]\) Bivariate Bicycle qLDPC code. The BP computation is offloaded to DPUs, where integer-oriented normalized Min-Sum operations, layered message updates, and frequently accessed decoding state are executed and maintained close to memory.

    \item We organize the decoding workload at two levels of parallelism. Within each DPU, 11 tasklets divide the check-node update workload of a single syndrome instance and synchronize at color-group boundaries using shared decoding state in WRAM. Across DPUs, independent syndrome instances are processed concurrently to exploit large-scale syndrome-level parallelism.

    \item We evaluate the proposed scheme against a 16-logical-CPU baseline using projected aggregate throughput, per-syndrome processing time, logical error rate (LER), and single syndrome tail latency. The evaluated configuration achieves up to an \(8.8\times\) improvement in projected aggregate throughput while maintaining sub-millisecond decoder-kernel latency over the evaluated one-to-five-iteration range.
\end{itemize}

The remainder of this paper is organized as follows. Section 2 provides background on qLDPC BP decoding, DPU-based PIM architectures, and trapped-ion decoding requirements. Section 3 describes the proposed DPU-based mapping and workload partitioning. Section 4 presents the evaluation setup and results. Section 5 concludes the paper.

\section{Background}

\subsection{Quantum LDPC Codes and Belief Propagation Decoding}

Quantum Low-Density Parity-Check (qLDPC) codes have attracted significant attention as a promising approach to reducing the physical qubit overhead of large-scale fault-tolerant quantum computation~\cite{breuckmann2021quantum,tillich2014quantum,bravyi2024high}. Compared with conventional two-dimensional local topological codes such as the surface code, qLDPC codes can support higher encoding rates while maintaining sparse parity-check structures~\cite{fowler2012surface,breuckmann2021quantum,bravyi2024high}. In a qLDPC code, each physical qubit participates in a limited number of stabilizer checks, and each check acts on a limited number of qubits, resulting in a bounded-degree sparse connectivity structure~\cite{breuckmann2021quantum}.

\begin{figure}[htbp]
    \centering
    \includegraphics[width=0.85\linewidth]{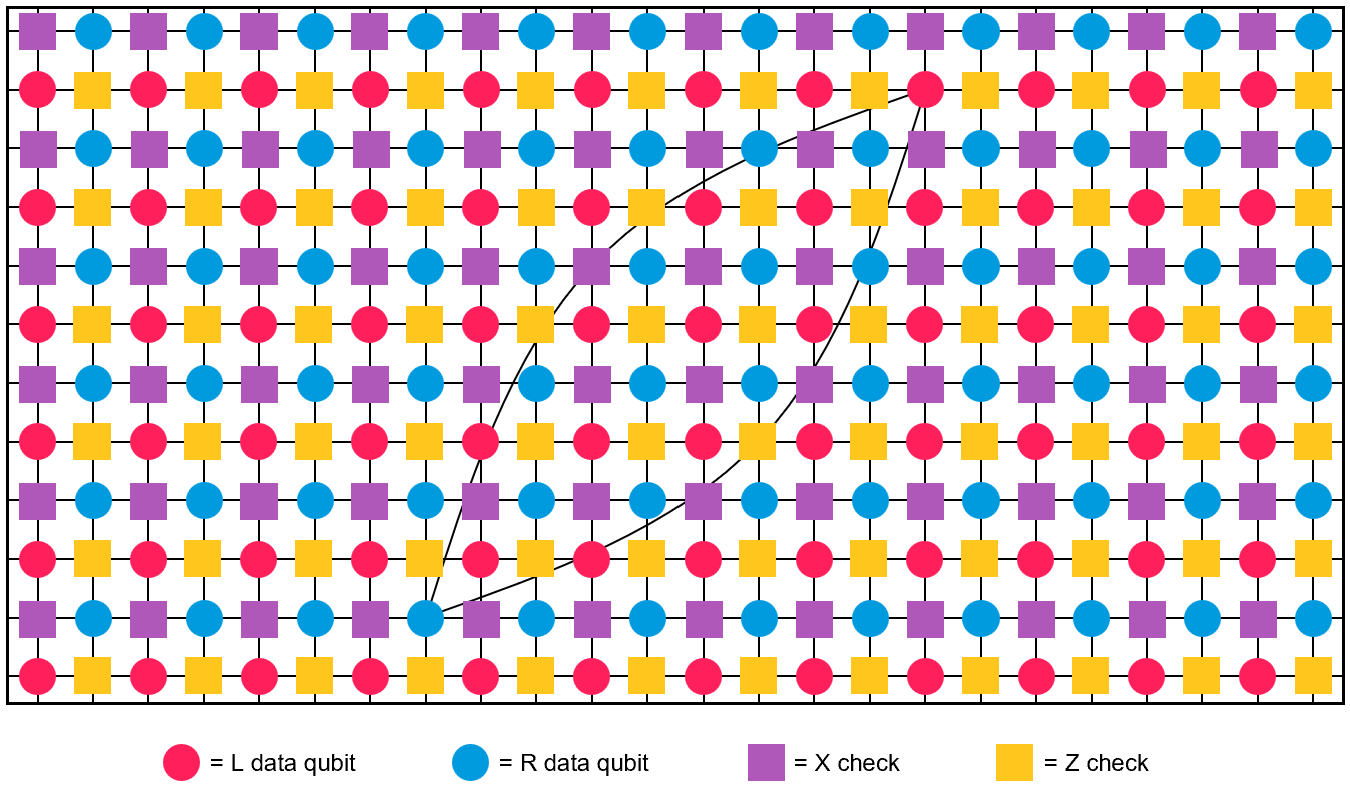}
    \caption{Tanner-graph structure of the \([[144,12,12]]\) Bivariate Bicycle qLDPC code used in this work. Circles represent data (variable) nodes, while squares represent check nodes. The data nodes are divided into two registers, and the check nodes are divided into two stabilizer-check types. Long-range connections are shown schematically, while individual edge types are not distinguished for visual clarity.}
    \label{fig:qLDPC_Tanner}
\end{figure}

The parity-check structure of a qLDPC code can be represented as a Tanner graph~\cite{Tanner1981recursive}. A Tanner graph is a bipartite graph composed of variable nodes and check nodes, where variable nodes represent physical qubits or error variables and check nodes represent stabilizer measurements or parity constraints. An edge indicates that the corresponding variable participates in a particular check. Figure~\ref{fig:qLDPC_Tanner} shows the Tanner-graph structure of the \([[144,12,12]]\) Bivariate Bicycle (BB) qLDPC code considered in this work~\cite{bravyi2024high}. The sparse connectivity of this graph limits each node update to information from a small set of neighboring nodes, providing a natural structure for iterative message-passing decoding~\cite{kschischang2001factor,poulin2008iterative}.

Belief Propagation (BP) exploits this sparse Tanner-graph representation by iteratively exchanging reliability messages between neighboring variable and check nodes~\cite{kschischang2001factor, Tanner1981recursive, poulin2008iterative,pearl1988probabilistic}. The decoder begins by assigning each variable node an initial reliability based on the physical error probability, typically represented by a log-likelihood ratio (LLR)~\cite{mackay1999good,richardson2001design}. Variable nodes send reliability messages to their neighboring check nodes, and each check node combines the incoming messages with the measured syndrome to enforce the corresponding parity constraint~\cite{kschischang2001factor, mackay1999good}. The resulting check-to-variable messages are returned to the variable nodes and combined with the channel prior to update the posterior reliability. A hard decision is then obtained from the updated reliability values, and the message-passing procedure is repeated over successive BP iterations~\cite{mackay1999good, richardson2001design}.

The sparse Tanner graph makes individual BP updates dependent only on neighboring nodes and therefore exposes fine-grained parallelism in the message-update process~\cite{breuckmann2021quantum, poulin2008iterative}. At the same time, iterative BP decoding retains edge-associated messages and node reliabilities and repeatedly accesses and updates this state over successive iterations. The resulting workload combines sparse arithmetic with repeated memory accesses, and its processing cost increases with the number of Tanner-graph edges and BP iterations. In a fault-tolerant system where syndrome instances are generated and decoded continuously, the same message-passing workload is executed repeatedly across many independent syndrome instances. These characteristics make memory locality and parallel processing capacity important architectural considerations for high-throughput qLDPC BP decoding.

\subsection{Near-Memory Processing and DPU-Based PIM Architecture}

Iterative BP decoding repeatedly accesses message state and node reliabilities while exposing parallelism across local message updates and independent syndrome instances. For such workloads, frequent transfers through a conventional processor-memory hierarchy can make data movement an important component of execution cost~\cite{mutlu2019processing,wulf1995hitting}. Near-Memory Processing (NMP) provides an architectural approach in which computation is performed close to memory, allowing frequently accessed data to remain near the processing units~\cite{asifuzzaman2023survey}. In this work, near-memory processing refers to this execution principle, while DPU-based Processing-in-Memory (PIM) denotes the concrete architecture used to realize it. UPMEM-PIM provides a programmable PIM platform in which lightweight processing cores, referred to as Data Processing Units (DPUs), are integrated with DRAM and execute kernels using DPU-local memory resources~\cite{gomezluna2022benchmarking,hyun2024pathfinding,devaux2019true}.

\begin{figure}[htbp]
    \centering
    \includegraphics[width=0.85\linewidth]{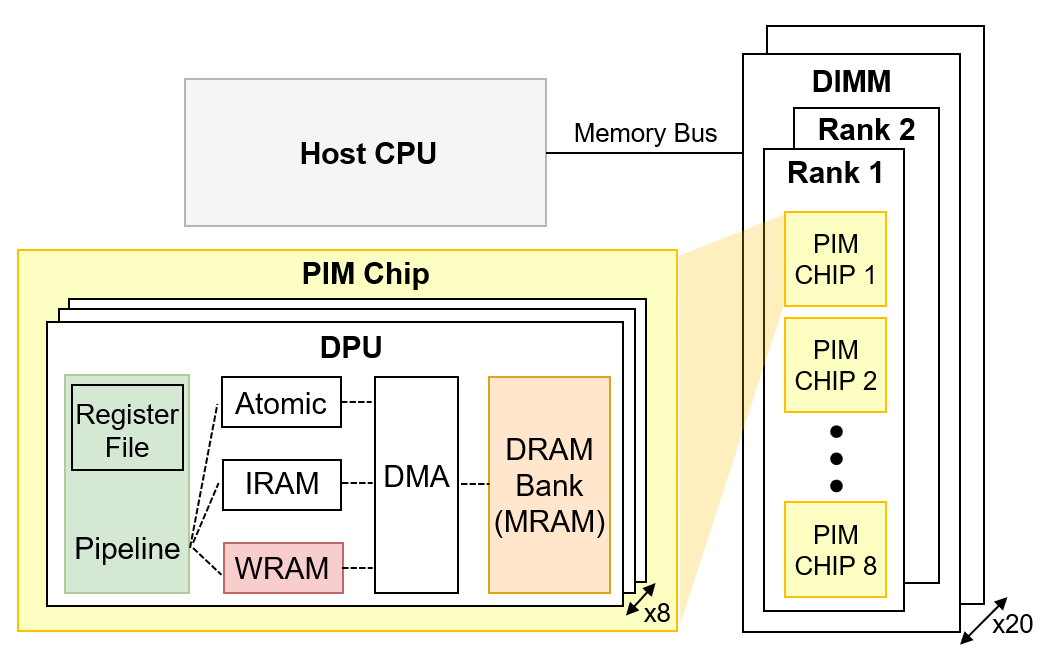}
    \caption{DPU-based PIM architecture used as the near-memory processing platform in this work. The host CPU communicates with PIM-enabled DIMMs through the memory bus. Each DIMM contains two ranks, each rank contains eight PIM chips, and each PIM chip contains eight DPUs, giving 2,560 DPUs in the 20-DIMM configuration. Each DPU includes a processing pipeline, register file, atomic unit, IRAM, WRAM, DMA support, and a local DRAM bank referred to as MRAM.}
    \label{fig:dpu_arch}
\end{figure}

Figure~\ref{fig:dpu_arch} illustrates the DPU-based PIM architecture considered in this work. The host CPU performs application-level control and transfers programs and data to PIM-enabled DIMMs through the memory interface, while the DPUs execute data in parallel kernels~\cite{gomezluna2022benchmarking,hyun2024pathfinding,friesel2023fullsystem}. The system configuration consists of 20 double-ranked PIM DIMMs. Each rank contains eight PIM chips, and each PIM chip contains eight DPUs, yielding a total of \(20 \times 2 \times 8 \times 8 = 2,560\) DPUs~\cite{gomezluna2022benchmarking, baumstark2023accelerating, hyun2024pathfinding, friesel2023fullsystem}. Independent data partitions can therefore be processed concurrently across a large number of DPUs, providing system-level parallelism.

Each DPU contains an in-order processing pipeline and several local memory structures~\cite{gomezluna2022benchmarking, baumstark2023accelerating, hyun2024pathfinding, shi2025dimm}. IRAM stores instructions, WRAM provides software-managed storage for frequently accessed working data, and MRAM provides a larger local DRAM space for input and output data. Data are transferred between MRAM and WRAM through DMA operations~\cite{gomezluna2022benchmarking, baumstark2023accelerating, hyun2024pathfinding}. A DPU supports up to 24 hardware threads, referred to as tasklets, which share the processing pipeline and local memory resources~\cite{gomezluna2022benchmarking, hyun2024pathfinding}. Instructions from different tasklets are interleaved through fine-grained multithreading, and consecutive instructions from the same tasklet follow the 11-cycle revolver scheduling constraint~\cite{hyun2024pathfinding}. The number of active tasklets and the distribution of work among them therefore influence utilization of the shared DPU pipeline.

The local memory hierarchy and large-scale DPU parallelism are well aligned with the repeated data accesses and independent syndrome-level workloads of qLDPC BP decoding. These architectural properties provide the basis for near-memory execution aimed at high-throughput BP decoding.

\subsection{Real-Time Decoding Requirements in Trapped-Ion Platforms}

Real-time quantum error correction requires the classical decoder to process syndrome information at a rate compatible with the QEC cycle and to provide correction information or Pauli-frame updates within the timing requirements of fault-tolerant control~\cite{battistel2023realtime,ryananderson2021realization}. The available decoding time depends on the underlying qubit technology and control architecture. Trapped-ion platforms generally operate with longer gate and measurement timescales than superconducting platforms and are commonly associated with a correspondingly longer decoder timing budget~\cite{haffner2008quantum,bruzewicz2019Trapped}. This millisecond-scale regime provides a relevant timing context for evaluating high-throughput qLDPC decoding.

Recent decoder studies provide concrete references for this timing regime. AlphaQubit identifies a target decoding time of approximately \(1~\mathrm{ms}\) per QEC round for trapped-ion devices, compared with approximately \(1~\mu\mathrm{s}\) per round for superconducting qubits~\cite{bausch2024learning}. More directly for qLDPC decoding, the beam-search decoder for the \([[144,12,12]]\) Bivariate Bicycle code evaluates the 99.9th-percentile runtime per-syndrome extraction round and identifies a configuration with a runtime below \(1~\mathrm{ms}\) as a promising candidate for trapped-ion architectures~\cite{ye2025beam}. Based on these studies, \(1~\mathrm{ms}\) is adopted in this work as a decoder-side timing reference for evaluating the BP compute stage. The reported DPU latency represents the serialized \(X+Z\) decoder-kernel computation associated with a single syndrome instance, providing a consistent reference for comparing decoder execution time with the millisecond-scale timing regime.

Real-time operation also depends on the distribution of decoding completion times. Occasional long-running decoding instances can delay the availability of correction information and contribute to a backlog of pending syndrome data even when typical execution times remain within the target range~\cite{barroso2013tail,battistel2023realtime}. Tail latency therefore provides an important complement to aggregate throughput and average processing time when evaluating a decoder for real-time QEC. Percentile-based latency and the maximum sampled completion time are consequently useful for characterizing whether decoder execution remains predictable under repeated syndrome processing.

\section{Proposed Method}

\begin{figure}[htbp]
    \centering
    \includegraphics[width=0.85\linewidth]{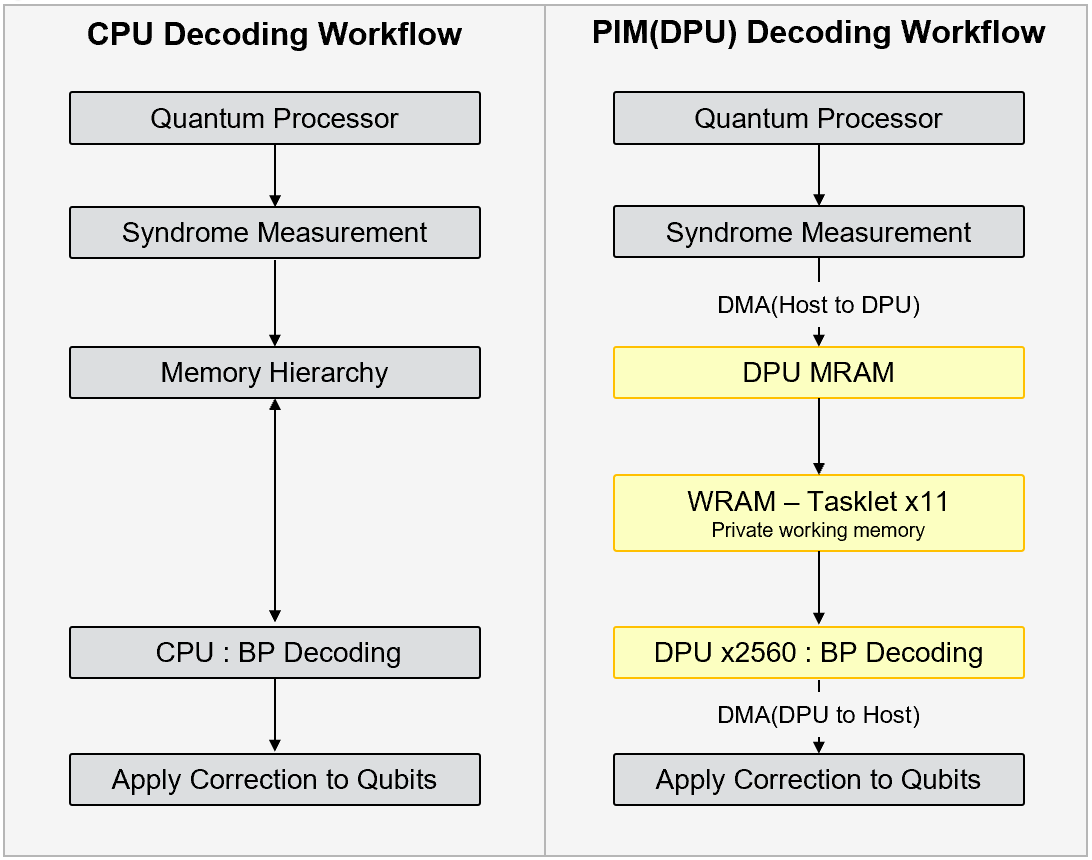}
    \caption{Overview of the conventional CPU-based and proposed DPU-based PIM decoding workflows. The proposed workflow offloads normalized Min-Sum BP computation to the DPUs and distributes independent syndrome instances across multiple DPUs for parallel decoding.}
    \label{fig:cpu_pim_workflow}
\end{figure}

To exploit the memory-access characteristics and parallelism of qLDPC BP decoding, we propose a near-memory execution scheme that offloads the normalized Min-Sum BP computation from the host CPU to a DPU-based Processing-in-Memory (PIM) architecture. Figure~\ref{fig:cpu_pim_workflow} provides an overview of the conventional CPU-based and proposed DPU-based decoding workflows. In the proposed workflow, syndrome data are transferred to the DPUs and processed using DPU-local memory. During each decoding pass, each DPU processes one syndrome instance, while independent syndrome instances are distributed across multiple DPUs to provide system-level parallelism for high aggregate decoding throughput. Section~\ref{sec:Normalized Min-Sum BP for DPU Execution} describes the normalized Min-Sum BP formulation for DPU execution, Section~\ref{sec:Intra-DPU Workload Partitioning and Memory Organization} presents the intra-DPU workload partitioning and memory organization.

\subsection{~\label{sec:Normalized Min-Sum BP for DPU Execution}Normalized Min-Sum BP for DPU Execution}

Standard log-domain Belief Propagation (BP) computes check-to-variable messages using nonlinear functions such as \(\tanh\) and \(\mathrm{atanh}\)~\cite{kschischang2001factor,mackay1999good,gallager1962low}. Normalized Min-Sum BP provides a reduced-complexity formulation of BP by replacing the nonlinear magnitude computation in the check-node update with sign evaluation, minimum-magnitude selection, and normalization~\cite{fossorier1999reduced,chen2002near,chen2002density,chen2005implementation,myung2017offset}. Its primary check-node operations can consequently be expressed using comparisons, additions, sign operations, and fixed-point scaling. The UPMEM-based DPU architecture used in this work employs an integer-oriented execution pipeline, while floating-point operations are supported through software routines~\cite{gomezluna2022benchmarking,hyun2024pathfinding}. This execution model is well aligned with the arithmetic structure of normalized Min-Sum BP, allowing its reduced-complexity message updates to be implemented primarily with integer operations supported by the DPU. Based on this combination of algorithmic efficiency and architectural compatibility, normalized Min-Sum BP is used as the decoding algorithm in the proposed DPU-based execution scheme.

\begin{figure}[htbp]
    \centering
    \includegraphics[width=0.85\linewidth]{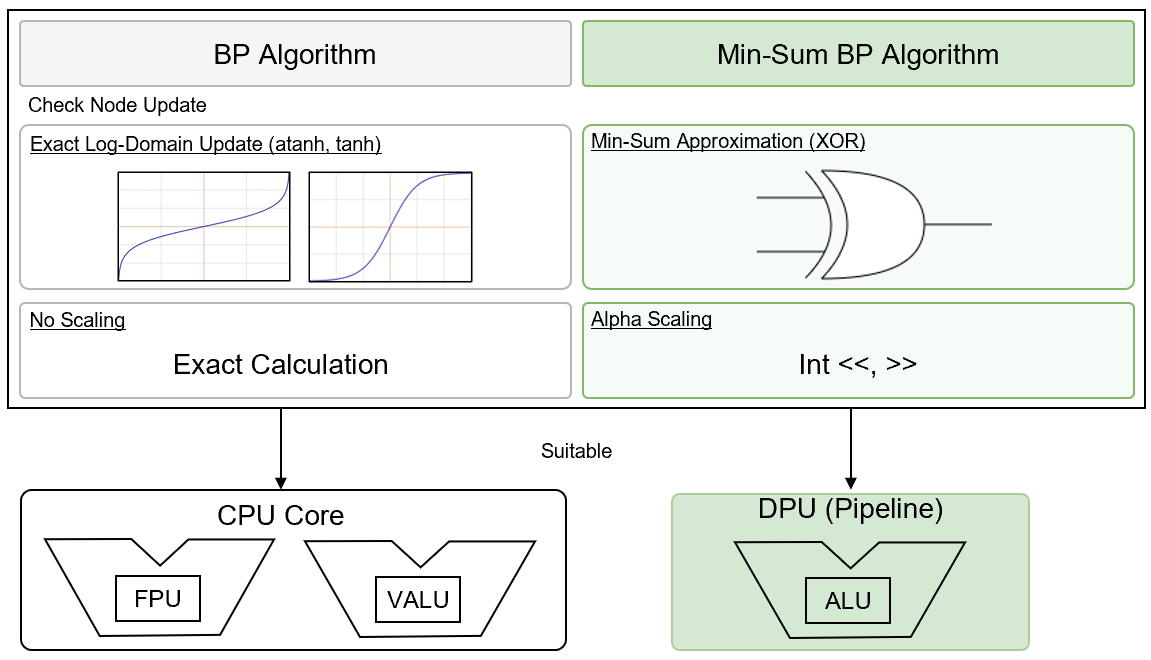}
    \caption{Arithmetic comparison of standard log-domain BP and normalized Min-Sum BP for DPU execution. Standard BP evaluates nonlinear functions such as \(\tanh\) and \(\mathrm{atanh}\) in the check-node update. Normalized Min-Sum uses sign evaluation, minimum-magnitude selection, and fixed-point scaling. The evaluated implementation uses \(\alpha=0.75=3/4\), allowing the normalization step to be expressed using integer addition and bit shifting.}
    \label{fig:minsum_dpu_fit}
\end{figure}

Figure~\ref{fig:minsum_dpu_fit} compares the arithmetic structures of standard log-domain BP and normalized Min-Sum BP from the DPU execution perspective. The normalized Min-Sum implementation used in this work fixes the normalization factor to \(\alpha=0.75=3/4\). With this value, the normalization step in the fixed-point implementation can be expressed using integer addition and bit shifting.

Let \(m_{v\rightarrow c}\) denote the message from variable node \(v\) to check node \(c\), and let \(m_{c\rightarrow v}\) denote the message in the opposite direction. Let \(N(c)\) and \(N(v)\) denote the neighboring variable and check nodes of \(c\) and \(v\), respectively. For a measured syndrome value \(s_c\), the normalized Min-Sum check-node update is expressed as
\[
m_{c\rightarrow v}^{\mathrm{new}}
=
\alpha
(-1)^{s_c}
\prod_{u\in N(c)\setminus{v}}
\operatorname{sign}(m_{u\rightarrow c})
\min_{u\in N(c)\setminus{v}}
\left|m_{u\rightarrow c}\right|.
\]
The sign of the outgoing message is determined by the syndrome value and the signs of the incoming messages. Its magnitude is obtained from the minimum incoming magnitude excluding the target variable node and is then scaled by the normalization factor \(\alpha\). The normalization factor compensates for the tendency of the basic Min-Sum approximation to overestimate check-node reliability~\cite{fossorier1999reduced,chen2002near,chen2002density,chen2005improved,myung2017offset}. For the fixed value \(\alpha=3/4\), scaling a non-negative magnitude (a) is implemented as \(((a\ll 1)+a)\gg 2\), before the message sign is applied. The normalization operation is therefore represented using integer addition and bit-shift operations in the DPU kernel.

The variable-node update combines the channel prior \(L_v\) with incoming check-to-variable messages:
\[
m_{v\to c}
=
L_v
+
\sum_{c'\in N(v)\setminus\{c\}}
m_{c'\to v}.
\]
The decoder uses a row-layered normalized Min-Sum schedule with a running posterior reliability \(\lambda_v\). At initialization,
\[
\lambda_v=L_v,
\qquad
m_{c\rightarrow v}=0.
\]
Before check node \(c\) is updated, its previous contribution is removed from the running posterior:
\[
m_{v\rightarrow c}
=
\lambda_v-m_{c\rightarrow v}^{\mathrm{old}}.
\]
After \(m_{c\rightarrow v}^{\mathrm{new}}\) is computed, the check-to-variable message and running posterior are updated in place:
\[
m_{c\rightarrow v}
\leftarrow
m_{c\rightarrow v}^{\mathrm{new}},
\qquad
\lambda_v
\leftarrow
m_{v\rightarrow c}+m_{c\rightarrow v}^{\mathrm{new}}.
\]
The row-layered schedule immediately incorporates each newly computed check-node message into the running posterior, allowing the updated reliability information to be used by subsequent check-node updates within the same BP iteration. In the proposed implementation, the message state and posterior reliability are therefore updated in place throughout the iterative decoding process.

The check-to-variable messages and running posterior values are represented as signed 32-bit integers during decoding. After the final BP iteration, the posterior value is clipped as
\[
L_v^{\mathrm{post}}
=
\operatorname{clip}_{[-32000,\,32000]}(\lambda_v),
\]
and stored as a signed 16-bit integer. The estimated error component is obtained from the sign of the posterior reliability:
\[
\hat{e}_v
=
\begin{cases}
1, & L_v^{\mathrm{post}}\leq 0,\\
0, & L_v^{\mathrm{post}}>0.
\end{cases}
\]
These design choices provide an arithmetic-level match between normalized Min-Sum BP and the UPMEM-based DPU execution model. The check-node update is expressed through sign operations, minimum selection, addition, and bit shifting. The layered schedule updates reliability information in place, and the iterative decoding state is represented using integer data types. Together, these properties form the computational basis for executing normalized Min-Sum BP on the DPU architecture.

\subsection{~\label{sec:Intra-DPU Workload Partitioning and Memory Organization}Intra-DPU Workload Partitioning and Memory Organization}

\begin{figure}[htbp]
    \centering
    \includegraphics[width=0.85\linewidth]{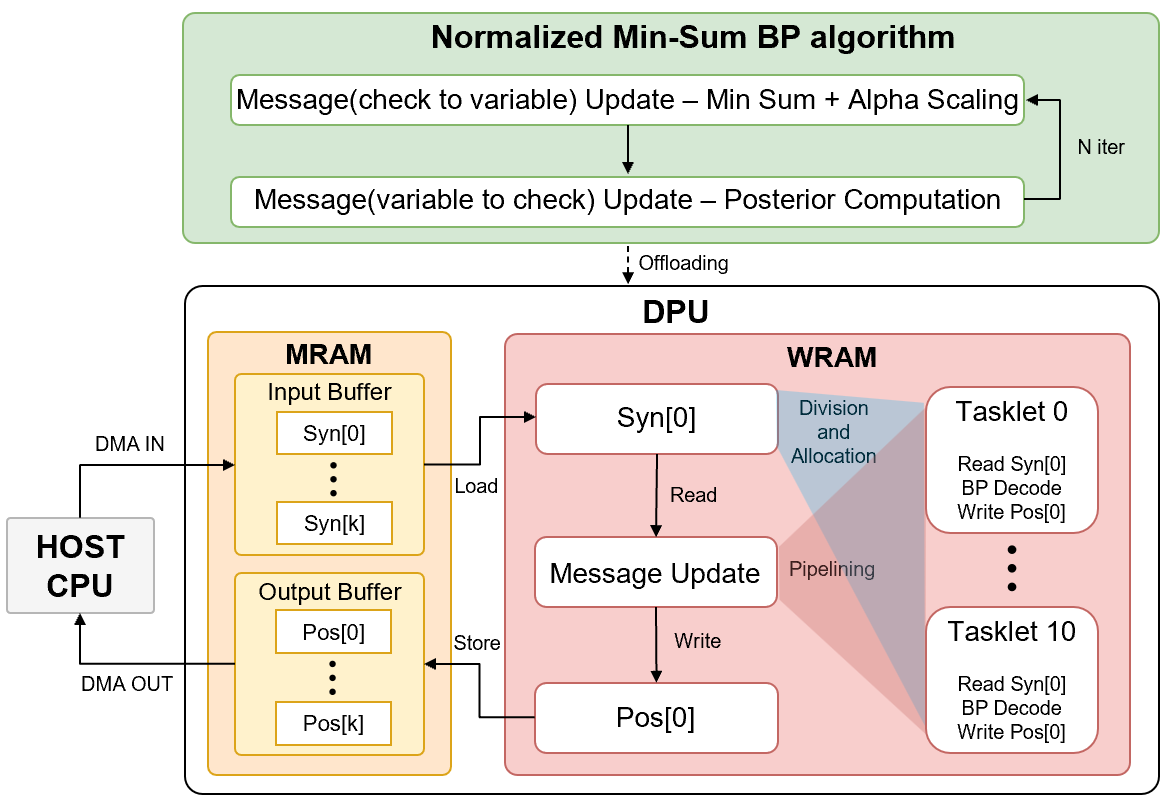}
    \caption{Processing flow of the proposed normalized Min-Sum BP decoder within a DPU. For each \(X\)- or \(Z\)-error decoding, one syndrome instance is transferred through MRAM and processed using decoding data stored in WRAM. The check-node computations are divided among 11 tasklets during iterative BP decoding, and the posterior output is stored in MRAM after the specified number of iterations.}
    \label{fig:dpu_bp_overview}
\end{figure}

For each \(X\)- or \(Z\)-error decoding, one DPU processes one syndrome instance. Figure~\ref{fig:dpu_bp_overview} shows the overall processing flow within a DPU. The host CPU transfers the syndrome data to DPU MRAM, and the working data required for iterative normalized Min-Sum BP decoding are then loaded into WRAM. During decoding, the message values and posterior reliabilities stored in WRAM are repeatedly updated over the specified number of BP iterations. After the iterations are completed, the posterior output is stored in MRAM and transferred back to the host.

The memory organization follows the repeated data access pattern of iterative BP decoding. MRAM is used mainly for data transfer between the host and the DPU, while the syndrome, check-to-variable messages, and running posterior reliabilities are maintained in WRAM during decoding. These values are repeatedly read and updated by the row-layered normalized Min-Sum schedule. Keeping this intermediate state in DPU-local WRAM allows the decoder to reuse the same data throughout successive BP iterations. Data transfers between MRAM and WRAM are performed through DMA operations, following the local-memory organization of UPMEM-based DPUs~\cite{gomezluna2022benchmarking, hyun2024pathfinding, friesel2023fullsystem}.

With the repeatedly accessed decoding state maintained in WRAM, the check-node computations for one syndrome are divided among 11 tasklets within the DPU. The tasklets use the same DPU processing pipeline and access the shared decoding state in WRAM while processing different subsets of check nodes. In the UPMEM-based DPU architecture, instructions from different tasklets are interleaved through fine-grained multithreading, and consecutive instructions from the same tasklet are separated by the 11-cycle revolver scheduling constraint~\cite{gomezluna2022benchmarking, hyun2024pathfinding}. The proposed implementation uses 11 tasklets to partition the check-node computations of the same syndrome across the shared DPU pipeline.

Applying this tasklet-level parallelism to the row-layered update requires the shared-data dependencies to be handled explicitly. Processing a check node \(c\) updates the running posterior reliability \(\lambda_v\) of each variable node connected to that check. If two check nodes share a variable node, simultaneous processing would cause both checks to update the same \(\lambda_v\). We therefore define two check nodes \(c_i\) and \(c_j\) as conflicting when
\[
N(c_i)\cap N(c_j)\neq\varnothing.
\]
For the \([[144,12,12]]\) Bivariate Bicycle qLDPC code used in this work, the 72 check nodes are divided into three conflict-free color groups,
\(\mathcal{G}_1\), \(\mathcal{G}_2\), and \(\mathcal{G}_3\). Check nodes assigned to the same group have non-overlapping variable-node neighborhoods:
\[
N(c_i)\cap N(c_j)=\varnothing,
\qquad
c_i,c_j\in\mathcal{G}_g,\quad i\neq j.
\]

Each variable node in this Tanner graph is connected to three check nodes. The three check nodes connected to the same variable therefore belong to different color groups so that no two checks within a group share that variable. The three-group partition used in the proposed decoder satisfies this condition for the fixed Tanner graph. As a result, check nodes within the same group update different posterior values and can be processed concurrently without write conflicts.

During decoding, the check nodes in one color group are distributed across the 11 tasklets. Each tasklet performs normalized Min-Sum updates for its assigned check nodes using the message and posterior data stored in WRAM. After all tasklets complete the processing of the group, a barrier synchronizes them before the next group begins. The next color group then uses the posterior reliabilities updated by the preceding group. In this way, check nodes within each color group are processed in parallel, while synchronization between groups preserves the update order of the row-layered normalized Min-Sum schedule.

Beyond the workload partitioning within each DPU, independent syndrome instances are distributed across multiple DPUs. In the 2,560-DPU configuration considered in this work, each DPU processes one syndrome instance, allowing independent decoding workloads to proceed concurrently without cross-DPU synchronization during BP computation~\cite{gomezluna2022benchmarking, hyun2024pathfinding}. This DPU-level syndrome parallelism provides the system-scale processing capacity used to increase aggregate decoding throughput.

\section{Results}
This section presents the results of the proposed DPU-based near-memory decoding scheme in comparison with a 16-logical-CPU baseline. We first describe the evaluation setup, including the qLDPC code parameters, error model, CPU and DPU environments, and measurement procedure. We then evaluate decoding throughput, the trade-off between logical error rate (LER) and per-syndrome processing time, and single syndrome tail latency. The latency results are compared with the \(1~\mathrm{ms}\) decoder-side timing reference considered for trapped-ion QEC.

\subsection{System and Workload Configuration}
The CPU baseline is evaluated on an Amazon EC2 m4.4xlarge instance equipped with an Intel Xeon E5-2686 v4 processor running at \(2.30~\mathrm{GHz}\). The instance provides 16 logical CPUs, corresponding to eight physical cores with two hardware threads per core. The CPU decoder is compiled with GCC 11.4.0 in an Ubuntu Linux environment. DPU execution is evaluated using uPIMulator, which provides DPU logical cycle counts while modeling tasklet scheduling and the revolver pipeline constraint. The simulated cycle counts are converted to execution time using a \(350~\mathrm{MHz}\) DPU clock frequency~\cite{gomezluna2022benchmarking, hyun2024pathfinding}. The decoding experiments target the \([[144,12,12]]\) Bivariate Bicycle (BB) qLDPC code~\cite{bravyi2024high}. To characterize decoder behavior under different physical error conditions, five component-wise physical error probabilities are considered:
\[
p \in \{0.001,\,0.003,\,0.005,\,0.007,\,0.009\}.
\]

For each data qubit, the \(X\)- and \(Z\)-error components are introduced independently with probability \(p\). The resulting single qubit Pauli probabilities are
\[
\Pr(I)=(1-p)^2,\qquad
\Pr(X)=p(1-p),\qquad
\Pr(Z)=p(1-p),\qquad
\Pr(Y)=p^2.
\]
For each \(X\)- or \(Z\)-component error vector \(e\), the corresponding syndrome is determined from the parity-check matrix (H) as
\[
s=He\pmod 2.
\]
No measurement errors are added during syndrome extraction. The syndrome is determined by the generated data qubit error pattern and the parity-check structure. The \(X\)- and \(Z\)-error components are decoded separately, and one complete decoding trial consists of one \(X\)-component decode and one \(Z\)-component decode.
For each component, the estimated correction \(\hat{e}\) is combined with the generated error to obtain the residual error,
\[
r=e\oplus\hat{e}.
\]
A component decode is considered successful when the estimated correction reproduces the input syndrome and the residual error belongs to the corresponding stabilizer row space. A complete \(X+Z\) trial is recorded as a logical failure when either component decode fails. The logical error rate (LER) is evaluated for BP iteration counts from 1 to 10, with a fixed iteration count applied to all trials in each configuration.
For each physical error probability and iteration count, 10 independent runs are performed using different random seeds, with \(10^7\) complete \(X+Z\) trials in each run. This provides \(10^8\) trials per configuration and sufficient failure observations in the low-LER regime considered in this work. The LER is estimated from the observed fraction of logical failures. Since this estimate is obtained from a finite number of success-or-failure outcomes, Wilson 95\% confidence intervals are reported to quantify its statistical uncertainty.

CPU throughput is measured using 50,000 complete \(X+Z\) decoding trials for each physical error probability and BP iteration count. A fixed number of trials is used for every configuration to maintain the same workload size across throughput measurements. The 16 logical CPUs process independent syndrome instances in parallel. Each configuration is measured 10 times using different random seeds, and throughput is reported as the mean number of complete decoding trials processed per second.
For DPU throughput, uPIMulator provides the cycle count for one \(X\)- or \(Z\)-component decoding pass using the proposed 11-tasklet configuration. A complete decode consists of sequential \(X\)- and \(Z\)-component passes, and the corresponding processing time is obtained from the simulated cycle counts at \(350~\mathrm{MHz}\). Cycle counts are collected for 10 independently generated syndrome inputs. Because the syndrome workloads assigned to different DPUs use independent decoding state and require no inter-DPU synchronization during BP decoding, the 2,560-DPU configuration is represented as concurrent execution of independent DPU decoding tasks. The per-DPU cycle counts provided by uPIMulator are used to determine the system throughput under this concurrent-execution model~\cite{gomezluna2022benchmarking,hyun2024pathfinding, friesel2023fullsystem}. The reported DPU metrics characterize decoder-kernel execution, which is evaluated at the cycle level using uPIMulator. Host--DPU data transfer and runtime orchestration are treated as separate system stages, consistent with the full-system organization of UPMEM-based execution in which data movement between the host and PIM modules constitutes a distinct part of the execution path~\cite{friesel2023fullsystem}. This stage-separated evaluation provides a consistent basis for characterizing the computation and parallel processing capacity of the proposed decoding kernel.

Single syndrome tail latency is evaluated for BP iteration counts from 1 to 5. For the CPU baseline, \(10^7\) complete \(X+Z\) decoding times are collected for each configuration. This sample count provides approximately \(10^4\) observations within the upper \(0.1\%\) of the measured distribution, allowing the \(p99.9\) latency region to be characterized from a sufficiently populated set of observations. For the DPU, the latency distribution is obtained from \(10^4\) independently seeded uPIMulator runs. The serialized \(X+Z\) latency is calculated from the simulated DPU cycle counts. The distributions are summarized using the mean, standard deviation, \(p50\), \(p90\), \(p99\), \(p99.9\), and the maximum sampled latency.

Before collecting the performance results, the DPU decoding program executed in uPIMulator is verified against the CPU reference decoder. Both decoders receive the same syndrome input and use the same fixed-point channel prior, normalization factor, row-layered update order, and BP iteration count. Their posterior outputs are compared byte by byte, and all configurations reported in this work satisfy this output-equivalence check. Formal definitions of throughput, per-syndrome processing time, tail latency, and LER are provided in Appendix~\ref{sec:Metric Definitions}.

\subsection{Throughput Comparison}

\begin{figure}[htbp]
    \centering
    \includegraphics[width=1\linewidth]{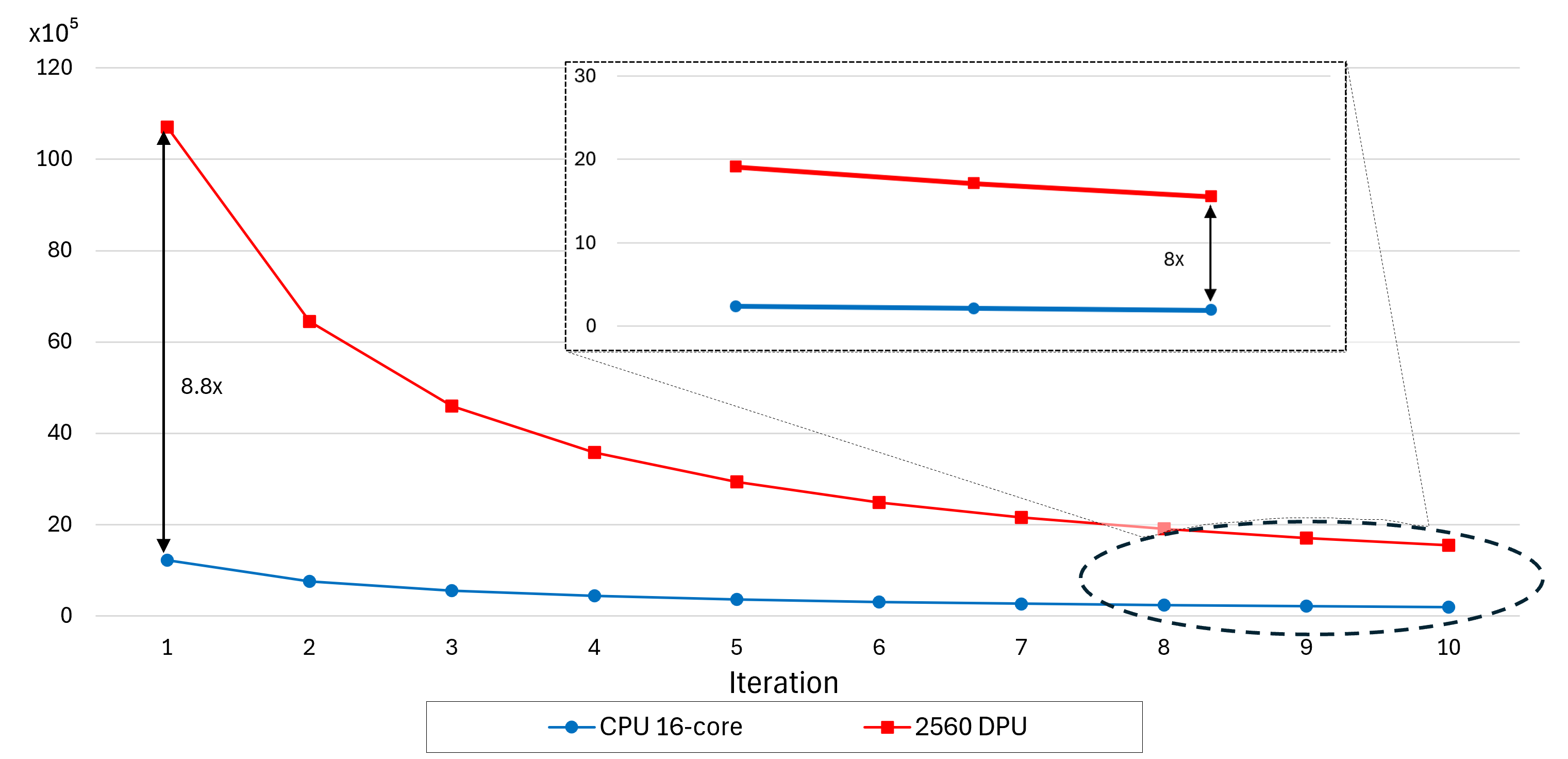}
    \caption{Decoding throughput of the 16-logical-CPU baseline and the 2,560-DPU configuration at \(p=0.001\) as the number of BP iterations increases. The DPU throughput is derived from per-DPU uPIMulator cycle counts under the concurrent independent-DPU execution model. The inset enlarges the throughput comparison for eight to ten BP iterations.}
    \label{fig:throughput_comparison}
\end{figure}

Figure~\ref{fig:throughput_comparison} compares the decoding throughput of the 16-logical-CPU baseline and the 2,560-DPU configuration at \(p=0.001\) as the number of BP iterations increases. The comparison represents the processing capacity of the two evaluated configurations. The CPU baseline processes independent syndrome instances across 16 logical CPUs, while the DPU configuration processes independent syndrome instances concurrently across 2,560 DPUs.

The 2,560-DPU configuration provides higher throughput than the CPU baseline across all evaluated iteration counts. At one BP iteration, the CPU baseline processes approximately \(1.22\times10^6\) complete decodes/s, while the 2,560-DPU configuration reaches approximately \(1.071\times10^7\) complete decodes/s, corresponding to an \(8.8\times\) difference. At ten iterations, the throughput decreases to approximately \(1.93\times10^5\) complete decodes/s for the CPU baseline and \(1.55\times10^6\) complete decodes/s for the DPU configuration, which corresponds to an \(8.0\times\) difference. Thus, the configuration-level throughput advantage is maintained as the BP workload increases with the iteration count.

The higher throughput of the DPU configuration is primarily associated with syndrome-instance-level parallelism across the DPU array. Each DPU uses 11 tasklets to process the check-node workload of one syndrome instance, while different DPUs process different syndrome instances independently. The 11 tasklets within a DPU therefore contribute to the execution of the same decoding task, whereas system-level throughput is increased by executing many independent decoding tasks concurrently across 2,560 DPUs. This workload organization matches the parallel execution characteristics reported for UPMEM systems, where independent data partitions can be distributed across multiple DPUs with minimal inter-DPU communication~\cite{gomezluna2022benchmarking, baumstark2023accelerating}. The throughput difference in Figure~\ref{fig:throughput_comparison} therefore reflects the large degree of syndrome-level parallelism available in the DPU configuration rather than a direct comparison between the computational performance of one DPU and one CPU core.

Throughput decreases on both configurations as the BP iteration count increases. Under the fixed-iteration row-layered schedule, each additional iteration repeats the check-node message calculations and variable-node reliability updates over the Tanner graph, increasing the amount of computation required for each syndrome~\cite{mackay1999good, richardson2001design}. This relationship is directly reflected in the DPU execution cost. At \(p=0.001\), the simulated cycle count for one component increases from 41,815 cycles at one iteration to 289,760 cycles at ten iterations, an increase of approximately \(6.9\times\). Over the same range, the 2,560-DPU throughput decreases from \(1.071\times10^7\) to \(1.55\times10^6\) complete decodes/s, also by approximately \(6.9\times\). The throughput reduction therefore follows the increase in the computation required to complete each syndrome as additional BP iterations are performed.

The physical error probability has a substantially smaller effect on throughput than the iteration count under the evaluated fixed-iteration execution. Changing \(p\) changes the generated error and syndrome patterns, while the decoder still traverses the same Tanner-graph structure and executes the same sequence of row-layered message-update stages for a fixed number of iterations. Consequently, the DPU cycle counts and throughput remain nearly constant across the evaluated physical error probabilities at a given iteration count. The CPU throughput also shows considerably smaller variation with \(p\) than with the BP iteration count. These results show that, under the fixed-iteration decoding scheme used in this work, the throughput is governed primarily by the number of BP iterations and the available syndrome-instance parallelism. The detailed numerical results supporting these observations are provided in Table~\ref{tab:full_results} of Appendix~\ref{sec:Full Simulation Results in Cooperative DPU Execution Mode}.

\subsection{Logical Error Rate and Per-Syndrome Processing Time Trade-off}



\begin{figure}[htbp]
    \centering
    \includegraphics[width=1\linewidth]{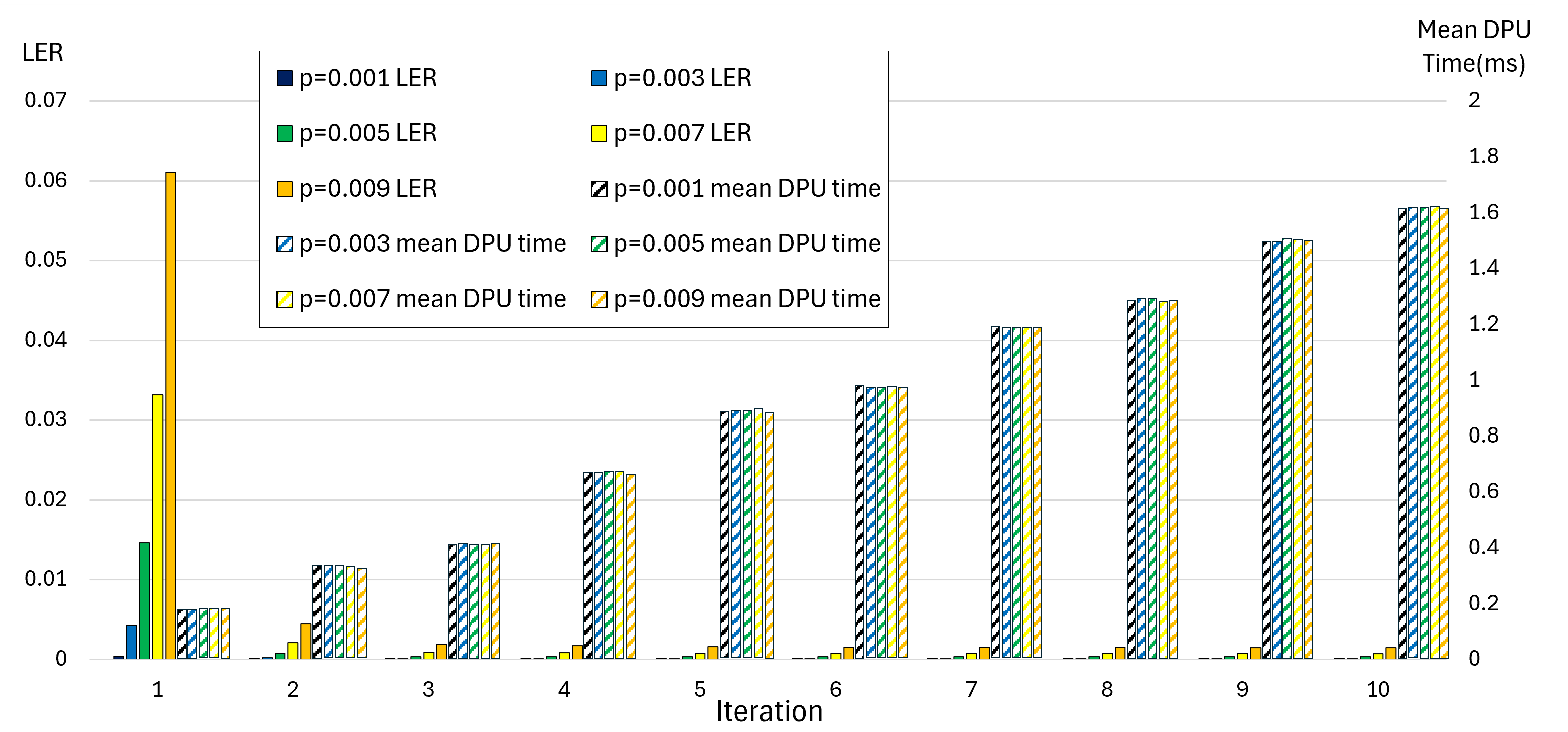}
    \caption{Logical error rate (LER) and mean per-syndrome DPU processing time as functions of the BP iteration count for the evaluated physical error probabilities. The processing time represents the serialized \(X+Z\) compute time required to decode one syndrome using the 11-tasklet DPU configuration.}
    \label{fig:latency_results}
\end{figure}

Figure~\ref{fig:latency_results} compares the logical error rate (LER) and per-syndrome DPU processing time as the number of BP iterations increases. The LER is evaluated for the five physical error probabilities considered in this work, while the processing time represents the compute time required to complete the serialized \(X\)- and \(Z\)-component decoding passes. The results exhibit a clear trade-off. Increasing the BP iteration count improves decoding performance, while each additional iteration increases the compute time required for one syndrome.

As a reference for decoding performance, we consider the break-even condition \(p_L(p)<p\), where \(p_L(p)\) denotes the measured LER at physical error probability \(p\). At one iteration, this condition is satisfied for \(p=0.001\), with an LER of \(3.628\times10^{-4}\). For the remaining physical error probabilities, the one-iteration LER remains above the corresponding value of \(p\). From two iterations onward, the condition is satisfied for all evaluated physical error probabilities.

The largest LER reduction occurs during the first few BP iterations. From one to two iterations, the LER decreases by approximately \(49.7\times\) at \(p=0.001\) and \(13.6\times\) at \(p=0.009\). The incremental improvement then becomes smaller: from two to three iterations, the LER decreases by approximately \(2.2\)--\(2.4\times\) across the evaluated values of \(p\). At \(p=0.001\), it reaches approximately \(3.1\times10^{-6}\) after three iterations and remains near this level, whereas at \(p=0.009\) it decreases more gradually from \(1.8829\times10^{-3}\) at three iterations to \(1.4334\times10^{-3}\) at ten iterations.

This behavior follows the iterative message-propagation and convergence characteristics of loopy BP~\cite{ihler2005loopy,murphy2013loopy}. During the early iterations, syndrome information is propagated through check-to-variable messages and repeatedly incorporated into the posterior reliabilities. Under the row-layered schedule, updated information from preceding layers is also available to subsequent layers within the same iteration. These early message-passing rounds can therefore produce substantial changes in variable reliabilities and hard decisions~\cite{mackay1999good, richardson2001design}. As the iterations continue, the posterior reliabilities of converging instances become progressively more stable, reducing the effect of subsequent message updates. The remaining failures also increasingly consist of error configurations that are difficult for standard BP to resolve through additional iterations alone. In qLDPC codes, short cycles and quantum degeneracy can interfere with BP convergence~\cite{poulin2008iterative,raveendran2021trapping,kuo2022exploiting}, which is consistent with the diminishing LER improvement observed after the first few iterations.

In contrast to the diminishing LER improvement, the per-syndrome DPU processing time increases almost linearly with the BP iteration count. The processing time is approximately \(0.239~\mathrm{ms}\) at one iteration, \(0.397~\mathrm{ms}\) at two iterations, \(0.556~\mathrm{ms}\) at three iterations, and \(0.872~\mathrm{ms}\) at five iterations, reaching approximately \(1.656~\mathrm{ms}\) at ten iterations. Consecutive iteration settings add approximately \(0.157\)--\(0.159~\mathrm{ms}\) of complete \(X+Z\) processing time. This behavior follows directly from the fixed-iteration execution structure. Early termination is not applied, and each additional iteration repeats the same row-layered check-node calculations, reliability updates, and color-group synchronization sequence over the fixed Tanner graph. The amount of additional computation introduced by each iteration is therefore nearly constant, producing the approximately linear increase in processing time observed in Figure~\ref{fig:latency_results}.

Taken together, these results establish a clear trade-off between LER and per-syndrome processing time. Each additional BP iteration incurs an approximately constant compute-time cost, while the corresponding LER reduction becomes progressively smaller after the first few iterations. This trade-off is particularly pronounced at lower physical error probabilities, where most of the observed LER improvement is obtained with relatively few iterations. At higher physical error probabilities, additional iterations continue to provide measurable LER improvement over a wider iteration range, although the marginal improvement gradually decreases. From the processing time perspective, five iterations require approximately \(0.872~\mathrm{ms}\), while six iterations increase the value to approximately \(1.029~\mathrm{ms}\). The evaluated one-to-five-iteration range therefore remains within the \(1~\mathrm{ms}\) decoder-side compute-time reference considered in this work. These results characterize the BP iteration count as a direct parameter for balancing decoding performance and per-syndrome processing time. Detailed numerical results for the LER and per-syndrome DPU processing time are provided in Table~\ref{tab:full_results} of Appendix~\ref{sec:Full Simulation Results in Cooperative DPU Execution Mode}.

\subsection{Single Syndrome Tail Latency}

\begin{figure}[htbp]
    \centering
    \includegraphics[width=1\linewidth]{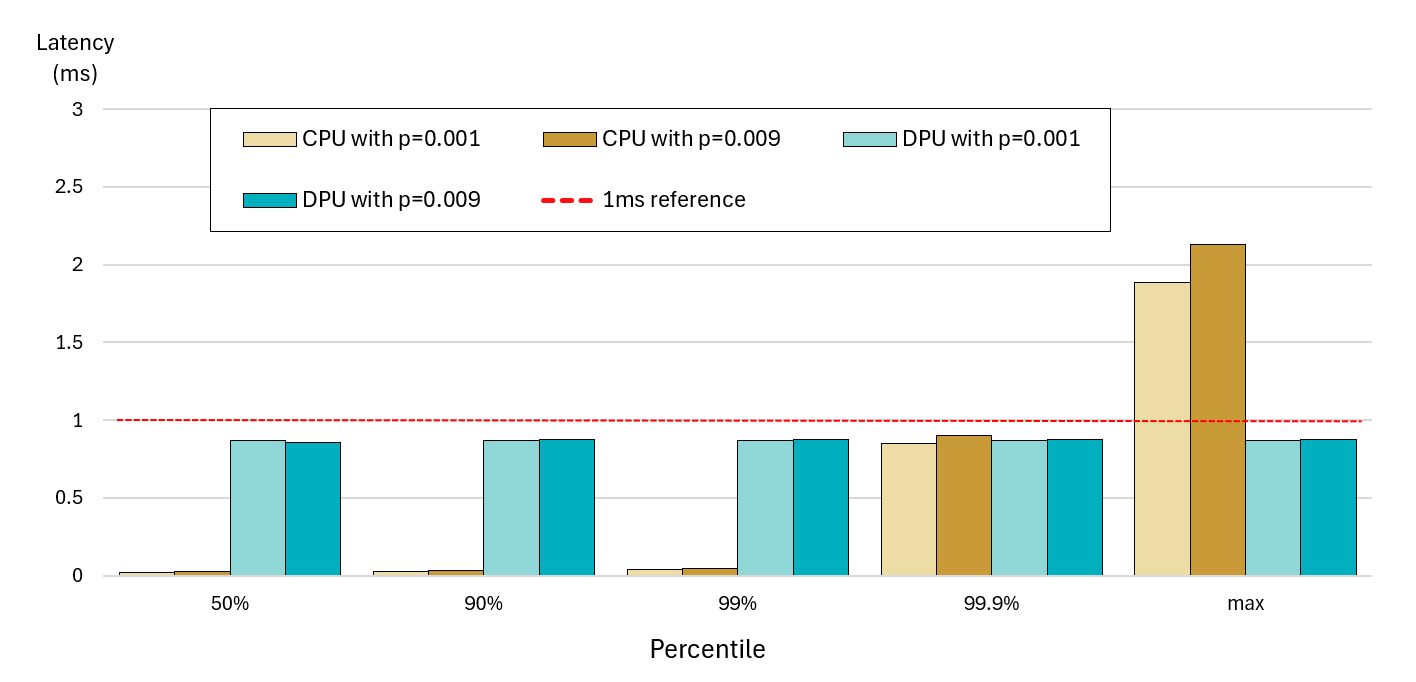}
    \caption{Comparison of the \(p50\), \(p90\), \(p99\), \(p99.9\), and maximum single syndrome latencies of the CPU baseline and DPU configuration at five BP iterations for p=0.001 and \(p=0.009\). The DPU values represent serialized \(X+Z\) compute-only latencies derived from uPIMulator cycle counts. The red dashed line indicates the \(1~\mathrm{ms}\) trapped-ion decoder-side latency reference.}
    \label{fig:tail_latency_comparison}
\end{figure}

Figure~\ref{fig:tail_latency_comparison} compares the single syndrome latency distributions of the CPU baseline and the DPU configuration at five BP iterations for \(p=0.001\) and \(p=0.009\). The distributions are characterized using \(p50\), \(p90\), \(p99\), \(p99.9\), and the maximum sampled latency, since high-percentile latency is important for identifying infrequent decoding delays that may not be visible from the average execution time alone~\cite{barroso2013tail}. The DPU values represent the serialized \(X+Z\) compute-only latency obtained from simulated DPU cycle counts.

The CPU baseline exhibits substantially lower latency over most of the distribution. As shown in Figure~\ref{fig:tail_latency_comparison}, the CPU \(p50\), \(p90\), and \(p99\) values remain far below the \(1~\mathrm{ms}\) reference for both physical error probabilities. The upper tail, however, increases sharply. The \(p99.9\) latency approaches the reference, and the maximum sampled latency exceeds \(1~\mathrm{ms}\) for both \(p=0.001\) and \(p=0.009\). This difference between typical and extreme execution times shows that a low median or average latency alone does not fully characterize the timing behavior relevant to real-time decoding.

The DPU latency distribution shows a different behavior. At five iterations, the simulated mean latency is \(869.260~\mu\mathrm{s}\) for \(p=0.001\) and \(857.161~\mu\mathrm{s}\) for \(p=0.009\). For \(p=0.001\), the median and the reported upper percentiles are approximately \(871.543~\mu\mathrm{s}\). For \(p=0.009\), the median is \(853.560~\mu\mathrm{s}\), while \(p90\) through the maximum remain at or below approximately \(872.606~\mu\mathrm{s}\). The corresponding standard deviations are \(5.843~\mu\mathrm{s}\) and \(7.759~\mu\mathrm{s}\), respectively, both below \(1\%\) of the mean latency. The distribution is therefore concentrated within a narrow timing range, and all reported percentiles and maximum sampled values remain below the \(1~\mathrm{ms}\) decoder-side reference.

The different tail behaviors are consistent with the execution characteristics of the two platforms. CPU BP decoding repeatedly accesses sparse Tanner-graph messages and posterior information through the CPU memory hierarchy, so its execution time can be affected by cache behavior, memory-bandwidth utilization, and data movement through the hierarchy. The CPU environment also includes system-level scheduling effects that can contribute to occasional long execution times. In the simulated DPU execution, the decoding kernel follows a fixed iteration count and a largely deterministic sequence of row-layered updates and color-group synchronization. The DPU pipeline modeled by uPIMulator also operates without conventional CPU caches or operating-system scheduling inside the DPU. These properties reduce input-dependent cycle-count variation in the evaluated simulation and are consistent with the narrow DPU latency distribution observed in Figure~\ref{fig:tail_latency_comparison}.

The five-iteration result represents the highest processing time within the one-to-five-iteration range emphasized in the tail-latency evaluation. Across all evaluated physical error probabilities and BP iteration counts from one to five, the maximum sampled serialized \(X+Z\) DPU compute latency remains below \(1~\mathrm{ms}\), with the largest observed value reaching approximately \(0.873~\mathrm{ms}\) at five iterations. Thus, within the evaluated simulation conditions, the DPU configuration maintains predictable sub-millisecond decoder-kernel latency throughout this iteration range. The reported DPU values characterize decoder-side compute time, which refers host--DPU communication and runtime-management stages are treated separately from this latency metric.

\section{Conclusion}

This work presents a DPU-based near-memory execution scheme for normalized Min-Sum BP decoding of the \([[144,12,12]]\) Bivariate Bicycle qLDPC code, in which 11 tasklets divide the row-layered workload of each syndrome while independent syndrome instances are processed concurrently across multiple DPUs. At \(p=0.001\), the 2,560-DPU configuration achieves approximately \(1.071\times10^7\) complete decodes/s at one BP iteration, corresponding to an \(8.8\times\) throughput difference relative to the 16-logical-CPU baseline. In the context of real-time decoding for trapped-ion platforms, the evaluated DPU configuration maintains the maximum sampled serialized \(X+Z\) compute latency below the \(1~\mathrm{ms}\) decoder-side reference over one to five BP iterations, with the largest sampled latency reaching approximately \(0.873~\mathrm{ms}\) at five iterations. At five iterations, the standard deviation of the sampled DPU latency is \(5.843~\mu\mathrm{s}\) at \(p=0.001\) and \(7.759~\mu\mathrm{s}\) at \(p=0.009\), corresponding to less than \(1\%\) of the mean latency in both cases. These results show that DPU-based near-memory processing can provide high syndrome-processing capacity together with predictable decoder-side compute latency for qLDPC BP decoding. The results further reveal a clear LER processing time trade-off. Most of the LER improvement is obtained during the first few BP iterations, while the per-syndrome processing time increases almost linearly with additional iterations, highlighting the importance of selecting the iteration count according to the required decoding performance and available real-time processing budget.


\section{Acknowledgements}
This work was supported by the National Research Foundation of Korea(NRF) grant funded by the Korea government(MSIT)(No. RS-2025-25434537).

\section{Data Availability}
The data are not deposited in a public repository, as no suitable field-specific repository is currently available for this type of simulation data. The data that support the findings of this study are available from the corresponding author upon reasonable request.


\clearpage
\appendix

\section{~\label{sec:Appendix}Appendix}
\subsection{~\label{sec:Metric Definitions}Metric Definitions}

This section defines the metrics used to evaluate the 16-logical-CPU baseline and the proposed DPU-based near-memory decoding scheme. The evaluation considers \textit{throughput}, \textit{per-syndrome DPU processing time}, \textit{single syndrome tail latency}, and \textit{logical error rate (LER)}. For the DPU-side throughput and timing metrics, host--DPU data transfer and runtime orchestration are treated as separate system stages.

\textit{Throughput} is defined as the number of complete \(X+Z\) decoding trials processed per second. For the CPU baseline, throughput is calculated as 
\[Throughput_{\mathrm{CPU}} = \frac{N_{\mathrm{decoded}}}{T_{\mathrm{exec}}},\] where \(N_{\mathrm{decoded}}\) is the number of complete decoding trials and \(T_{\mathrm{exec}}\) is the measured execution time required to process them. For the DPU evaluation, let \(C_{\mathrm{logic}}\) denote the uPIMulator-reported DPU logical cycle count for one \(X\)- or \(Z\)-component decoding pass, and let \(f_{\mathrm{DPU}}=350~\mathrm{MHz}\) denote the DPU clock frequency. The two components use the same kernel structure, data dimensions, and fixed iteration count and are processed as \(N_{\mathrm{pass}}=2\) serialized passes. The \textit{per-syndrome DPU processing time} is therefore
\[
{T_{\mathrm{syndrome}}}=
N_{\mathrm{pass}}
\frac{C_{\mathrm{logic}}}{f_{\mathrm{DPU}}}.
\]
Here, \(C_{\mathrm{logic}}\) is the DPU-wide cycle count for the 11-tasklet configuration, so the tasklet-level parallelism is already reflected in the measured cycle count. For \(N_{\mathrm{DPU}}=2560\), independent syndrome instances are assigned across the DPU array and processed concurrently. The corresponding throughput is calculated as
\[
{Throughput_{\mathrm{DPU}}}=
\frac{N_{\mathrm{DPU}}}{T_{\mathrm{syndrome}}}
=
\frac{N_{\mathrm{DPU}}f_{\mathrm{DPU}}}
{N_{\mathrm{pass}}C_{\mathrm{logic}}}.
\]
This calculation follows the concurrent independent-DPU execution model described in the main text and assumes balanced decoding workloads across the DPU array~\cite{gomezluna2022benchmarking, hyun2024pathfinding}. The per-DPU cycle counts obtained from uPIMulator are used to derive the throughput of the 2,560-DPU configuration under this execution model, characterizing its DPU-side decoder processing capacity.

\textit{Single syndrome tail latency} characterizes the high-percentile behavior of the decoding-time distribution. For the CPU baseline, each latency sample is obtained from the measured serialized \(X+Z\) decoding time of one complete trial. For the DPU, let \(C_{\mathrm{logic}, i}\) denote the logical cycle count obtained from independently seeded execution \(i\). Using the same \(N_{\mathrm{pass}}=2\), the corresponding DPU latency is
\[
N_{\mathrm{pass}}
\frac{C_{\mathrm{logic},i}}{f_{\mathrm{DPU}}}.
\]
The CPU and DPU latency distributions are summarized using the mean, standard deviation, \(p50\), \(p90\), \(p99\), \(p99.9\), and maximum sampled latency.

\textit{Logical rror rate (LER)} is defined as the fraction of complete \(X+Z\) decoding trials that result in a logical failure. For each error component, the decoder produces an estimated correction \(\hat{e}\) for the input syndrome associated with the generated error \(e\). The residual error is
\[
r=e\oplus\hat{e}.
\]
A component decode is considered successful when the estimated correction reproduces the input syndrome and the residual error belongs to the corresponding stabilizer row space. A complete \(X+Z\) trial is recorded as a failure when either component decode fails. The LER is calculated as
\[
\frac{N_{\mathrm{fail}}}{N_{\mathrm{trial}}},
\]
where \(N_{\mathrm{fail}}\) is the number of failed complete trials and \(N_{\mathrm{trial}}\) is the total number of complete trials. For each physical error probability and BP iteration setting, the Wilson 95\% confidence interval is calculated from the pooled failure count over the \(10^8\) complete trials used for the LER evaluation.

\subsection{~\label{sec:Full Simulation Results in Cooperative DPU Execution Mode}Full Simulation Results in Cooperative DPU Execution Mode}

Table~\ref{tab:full_results} presents the numerical results for all evaluated physical error probabilities and BP iteration counts. The table reports the DPU logical cycle count for one \(X\)- or \(Z\)-component decoding pass, the throughput of the 16-logical-CPU baseline and the 2,560-DPU configuration, the logical error rate (LER) with its Wilson 95\% confidence interval, and the per-syndrome DPU processing time. For each configuration, the CPU throughput is averaged over 10 measurements using different random seeds. The DPU logical cycle count and per-syndrome processing time are obtained from 10 independently generated syndrome inputs, and the throughput of the 2,560-DPU configuration is derived from the resulting per-DPU cycle count using the concurrent-execution model defined in Appendix~\ref{sec:Metric Definitions}. The LER and its Wilson 95\% confidence interval are calculated using the total number of failures observed over \(10^8\) complete \(X+Z\) decoding trials for each physical error probability and BP iteration setting.


\begingroup
\small
\setlength{\tabcolsep}{4pt}
\renewcommand{\arraystretch}{1.15}

\begin{longtable}{c|c|c|c|c|c|c|c}
\caption{Cooperative DPU simulation results averaged over 10 runs.}
\label{tab:full_results}\\
\hline
\(p\) & Iter. & \makecell[c]{DPU\\cycles}
& \makecell[c]{CPU thr.\\(dec/s)}
& \makecell[c]{DPU thr.\\(dec/s)}
& LER
& \makecell[c]{Wilson 95\% CI\\{[lo, hi]}}
& \makecell[c]{Per-Syndrome DPU time\\(\(\mu\mathrm{s}\))}
\\
\hline
\endfirsthead

\hline
\(p\) & Iter. & \makecell[c]{DPU\\cycles}
& \makecell[c]{CPU thr.\\(dec/s)}
& \makecell[c]{DPU thr.\\(dec/s)}
& LER
& \makecell[c]{Wilson 95\% CI\\{[lo, hi]}}
& \makecell[c]{Per-Syndrome DPU time\\(\(\mu\mathrm{s}\))}
\\
\hline
\endhead

\hline
\multicolumn{8}{r}{Continued on next page} \\
\hline
\endfoot

\hline
\endlastfoot

\multirow{10}{*}{0.001}
& 1 & 41,815 & 1,220,000 & 10,710,000
& 0.0003628
& \(\left[3.591,\,3.666\right]\times10^{-4}\)
& 238.943 \\ \cline{2-8}

& 2 & 69,498 & 762,200 & 6,450,000
& \(7.30\times10^{-6}\)
& \(\left[6.789,\,7.849\right]\times10^{-6}\)
& 397.131 \\ \cline{2-8}

& 3 & 97,360 & 556,300 & 4,600,000
& \(3.1\times10^{-6}\)
& \(\left[2.774,\,3.465\right]\times10^{-6}\)
& 556.343 \\ \cline{2-8}

& 4 & 125,072 & 440,900 & 3,580,000
& \(3.1\times10^{-6}\)
& \(\left[2.774,\,3.465\right]\times10^{-6}\)
& 714.697 \\ \cline{2-8}

& 5 & 152,520 & 359,800 & 2,940,000
& \(3.1\times10^{-6}\)
& \(\left[2.774,\,3.465\right]\times10^{-6}\)
& 871.543 \\ \cline{2-8}

& 6 & 179,968 & 309,100 & 2,490,000
& \(3.1\times10^{-6}\)
& \(\left[2.774,\,3.465\right]\times10^{-6}\)
& 1028.389 \\ \cline{2-8}

& 7 & 207,416 & 267,900 & 2,160,000
& \(3.1\times10^{-6}\)
& \(\left[2.774,\,3.465\right]\times10^{-6}\)
& 1185.234 \\ \cline{2-8}

& 8 & 234,864 & 236,900 & 1,910,000
& \(3.1\times10^{-6}\)
& \(\left[2.774,\,3.465\right]\times10^{-6}\)
& 1342.080 \\ \cline{2-8}

& 9 & 262,312 & 212,300 & 1,710,000
& \(3.1\times10^{-6}\)
& \(\left[2.774,\,3.465\right]\times10^{-6}\)
& 1498.926 \\ \cline{2-8}

& 10 & 289,760 & 193,000 & 1,550,000
& \(3.0\times10^{-6}\)
& \(\left[2.679,\,3.359\right]\times10^{-6}\)
& 1655.771 \\
\hline

\multirow{10}{*}{0.003}
& 1 & 41,815 & 1,210,000 & 10,710,000
& 0.0042847
& \(\left[4.272,\,4.298\right]\times10^{-3}\)
& 238.943 \\ \cline{2-8}

& 2 & 69,498 & 719,100 & 6,450,000
& \(1.684\times10^{-4}\)
& \(\left[1.659,\,1.710\right]\times10^{-4}\)
& 397.131 \\ \cline{2-8}

& 3 & 97,360 & 515,600 & 4,600,000
& \(7.080\times10^{-5}\)
& \(\left[6.917,\,7.247\right]\times10^{-5}\)
& 556.343 \\ \cline{2-8}

& 4 & 125,072 & 400,500 & 3,580,000
& \(6.77\times10^{-5}\)
& \(\left[6.611,\,6.933\right]\times10^{-5}\)
& 714.697 \\ \cline{2-8}

& 5 & 152,628 & 326,000 & 2,940,000
& \(6.67\times10^{-5}\)
& \(\left[6.512,\,6.832\right]\times10^{-5}\)
& 872.160 \\ \cline{2-8}

& 6 & 180,076 & 280,700 & 2,490,000
& \(6.61\times10^{-5}\)
& \(\left[6.453,\,6.771\right]\times10^{-5}\)
& 1029.006 \\ \cline{2-8}

& 7 & 207,524 & 248,000 & 2,160,000
& \(6.57\times10^{-5}\)
& \(\left[6.413,\,6.731\right]\times10^{-5}\)
& 1185.851 \\ \cline{2-8}

& 8 & 234,972 & 219,600 & 1,910,000
& \(6.56\times10^{-5}\)
& \(\left[6.403,\,6.721\right]\times10^{-5}\)
& 1342.697 \\ \cline{2-8}

& 9 & 262,420 & 198,800 & 1,710,000
& \(6.56\times10^{-5}\)
& \(\left[6.403,\,6.721\right]\times10^{-5}\)
& 1499.543 \\ \cline{2-8}

& 10 & 289,868 & 181,700 & 1,550,000
& \(6.35\times10^{-5}\)
& \(\left[6.196,\,6.508\right]\times10^{-5}\)
& 1656.389 \\
\hline
\clearpage

\multirow{10}{*}{0.005}
& 1 & 41,815 & 1,160,000 & 10,710,000
& 0.014621
& \(\left[1.460,\,1.464\right]\times10^{-2}\)
& 238.943 \\ \cline{2-8}

& 2 & 69,498 & 675,500 & 6,450,000
& 0.0007724
& \(\left[7.670,\,7.779\right]\times10^{-4}\)
& 397.131 \\ \cline{2-8}

& 3 & 97,360 & 471,700 & 4,600,000
& \(3.450\times10^{-4}\)
& \(\left[3.414,\,3.487\right]\times10^{-4}\)
& 556.343 \\ \cline{2-8}

& 4 & 125,072 & 371,200 & 3,580,000
& \(3.257\times10^{-4}\)
& \(\left[3.222,\,3.293\right]\times10^{-4}\)
& 714.697 \\ \cline{2-8}

& 5 & 152,628 & 300,600 & 2,940,000
& \(3.168\times10^{-4}\)
& \(\left[3.133,\,3.203\right]\times10^{-4}\)
& 872.160 \\ \cline{2-8}

& 6 & 180,076 & 261,800 & 2,490,000
& \(3.120\times10^{-4}\)
& \(\left[3.086,\,3.155\right]\times10^{-4}\)
& 1029.006 \\ \cline{2-8}

& 7 & 207,524 & 233,000 & 2,160,000
& \(3.115\times10^{-4}\)
& \(\left[3.081,\,3.150\right]\times10^{-4}\)
& 1185.851 \\ \cline{2-8}

& 8 & 234,972 & 209,400 & 1,910,000
& \(3.091\times10^{-4}\)
& \(\left[3.057,\,3.126\right]\times10^{-4}\)
& 1342.697 \\ \cline{2-8}

& 9 & 262,420 & 190,100 & 1,710,000
& \(3.081\times10^{-4}\)
& \(\left[3.047,\,3.116\right]\times10^{-4}\)
& 1499.543 \\ \cline{2-8}

& 10 & 289,868 & 174,500 & 1,550,000
& \(2.981\times10^{-4}\)
& \(\left[2.947,\,3.015\right]\times10^{-4}\)
& 1656.389 \\
\hline

\multirow{10}{*}{0.007}
& 1 & 41,815 & 1,160,000 & 10,710,000
& 0.033125
& \(\left[3.309,\,3.316\right]\times10^{-2}\)
& 238.943 \\ \cline{2-8}

& 2 & 69,498 & 647,500 & 6,450,000
& 0.0020749
& \(\left[2.066,\,2.084\right]\times10^{-3}\)
& 397.131 \\ \cline{2-8}

& 3 & 97,360 & 451,000 & 4,600,000
& 0.0008792
& \(\left[8.734,\,8.850\right]\times10^{-4}\)
& 556.343 \\ \cline{2-8}

& 4 & 125,072 & 342,000 & 3,580,000
& 0.0008054
& \(\left[7.999,\,8.110\right]\times10^{-4}\)
& 714.697 \\ \cline{2-8}

& 5 & 152,628 & 286,300 & 2,940,000
& \(7.721\times10^{-4}\)
& \(\left[7.667,\,7.776\right]\times10^{-4}\)
& 872.160 \\ \cline{2-8}

& 6 & 180,076 & 248,900 & 2,490,000
& \(7.510\times10^{-4}\)
& \(\left[7.456,\,7.564\right]\times10^{-4}\)
& 1029.006 \\ \cline{2-8}

& 7 & 207,524 & 221,300 & 2,160,000
& \(7.489\times10^{-4}\)
& \(\left[7.436,\,7.543\right]\times10^{-4}\)
& 1185.851 \\ \cline{2-8}

& 8 & 234,972 & 198,400 & 1,910,000
& \(7.402\times10^{-4}\)
& \(\left[7.349,\,7.455\right]\times10^{-4}\)
& 1342.697 \\ \cline{2-8}

& 9 & 262,420 & 182,500 & 1,710,000
& \(7.329\times10^{-4}\)
& \(\left[7.276,\,7.382\right]\times10^{-4}\)
& 1499.543 \\ \cline{2-8}

& 10 & 289,868 & 168,200 & 1,550,000
& \(7.108\times10^{-4}\)
& \(\left[7.056,\,7.160\right]\times10^{-4}\)
& 1656.389 \\
\hline

\multirow{10}{*}{0.009}
& 1 & 41,815 & 1,130,000 & 10,710,000
& 0.06107
& \(\left[6.102,\,6.112\right]\times10^{-2}\)
& 238.943 \\ \cline{2-8}

& 2 & 69,498 & 630,000 & 6,450,000
& 0.0044857
& \(\left[4.473,\,4.499\right]\times10^{-3}\)
& 397.131 \\ \cline{2-8}

& 3 & 97,360 & 439,900 & 4,600,000
& 0.0018829
& \(\left[1.874,\,1.891\right]\times10^{-3}\)
& 556.343 \\ \cline{2-8}

& 4 & 125,072 & 340,900 & 3,580,000
& 0.0016780
& \(\left[1.670,\,1.686\right]\times10^{-3}\)
& 714.697 \\ \cline{2-8}

& 5 & 152,706 & 276,700 & 2,930,000
& 0.0015832
& \(\left[1.575,\,1.591\right]\times10^{-3}\)
& 872.606 \\ \cline{2-8}

& 6 & 180,154 & 237,800 & 2,490,000
& 0.0015239
& \(\left[1.516,\,1.532\right]\times10^{-3}\)
& 1029.451 \\ \cline{2-8}

& 7 & 207,602 & 213,300 & 2,160,000
& 0.0015141
& \(\left[1.506,\,1.522\right]\times10^{-3}\)
& 1186.297 \\ \cline{2-8}

& 8 & 235,050 & 193,000 & 1,910,000
& 0.0014923
& \(\left[1.485,\,1.500\right]\times10^{-3}\)
& 1343.143 \\ \cline{2-8}

& 9 & 262,498 & 176,600 & 1,710,000
& 0.0014734
& \(\left[1.466,\,1.481\right]\times10^{-3}\)
& 1499.989 \\ \cline{2-8}

& 10 & 289,946 & 163,400 & 1,550,000
& 0.0014334
& \(\left[1.426,\,1.441\right]\times10^{-3}\)
& 1656.834 \\
\hline

\end{longtable}

\endgroup

\begin{thebibliography}{99}

\bibitem{terhal2015quantum}
Terhal B M, 2015, Quantum error correction for quantum memories, Reviews of Modern Physics, 87, 307--346

\bibitem{preskill1998reliable}
Preskill J, 1998, Reliable Quantum Computers, Proceedings of the Royal Society of London. Series A: Mathematical, Physical and Engineering Sciences, 454, 385--410

\bibitem{skoric2023parallel}
Skoric L, Hillmann T, Higgott O, Dilkes S and Browne D E, 2023, Parallel window decoding enables scalable fault tolerant quantum computation, Nature Communications, 14, 7040

\bibitem{breuckmann2021quantum}
Breuckmann N P and Eberhardt J N, 2021, Quantum Low-Density Parity-Check Codes, PRX Quantum, 2, 040101

\bibitem{tillich2014quantum}
Tillich J P and Zemor G, 2014, Quantum LDPC codes with positive rate and minimum distance proportional to the square root of the blocklength, IEEE Transactions on Information Theory, 60, 119--136

\bibitem{kschischang2001factor}
Kschischang F R, Frey B J and Loeliger H A, 2001, Factor graphs and the sum-product algorithm, IEEE Transactions on Information Theory, 47, 498--519

\bibitem{Tanner1981recursive}
Tanner R M, 1981, A Recursive Approach to Low Complexity Codes, IEEE Transactions on Information Theory, 27, 533--547

\bibitem{mutlu2019processing}
Mutlu O, 2019, Processing data where it makes sense: Enabling in-memory computation, Microprocessors and Microsystems, 67, 28--41

\bibitem{gomezluna2022benchmarking}
Gomez-Luna J, El Hajj I, Fernandez I, Giannoula C, Oliveira G F and Mutlu O, 2022, Benchmarking a New Paradigm: Experimental Analysis and Characterization of a Real Processing-in-Memory System, IEEE Access, 10, 52565--52608

\bibitem{das2022afs}
Das P, Pattison C A, Manne S, Carmean D M, Svore K M, Qureshi M and Delfosse N, 2022, AFS: Accurate, Fast, and Scalable Error-Decoding for Fault-Tolerant Quantum Computers, IEEE International Symposium on High-Performance Computer Architecture, 259--273

\bibitem{poulin2008iterative}
Poulin D and Chung Y, 2008, On the iterative decoding of sparse quantum codes, Quantum Information and Computation, 8, 987--1000

\bibitem{ye2025beam}
Ye M, Wecker D and Delfosse N, 2025, Beam search decoder for quantum LDPC codes, arXiv preprint, arXiv:2512.07057

\bibitem{fowler2012surface}
Fowler A G, Mariantoni M, Martinis J M and Cleland A N, 2012, Surface codes: Towards practical large-scale quantum computation, Physical Review A, 86, 032324

\bibitem{bravyi2024high}
Bravyi S, Cross A W, Gambetta J M, Maslov D, Rall P and Yoder T J, 2024, High-threshold and low-overhead fault-tolerant quantum memory, Nature, 627, 778--782

\bibitem{pearl1988probabilistic}
Pearl J, 1988, Probabilistic Reasoning in Intelligent Systems: Networks of Plausible Inference, Morgan Kaufmann

\bibitem{mackay1999good}
MacKay D J C, 1999, Good error-correcting codes based on very sparse matrices, IEEE Transactions on Information Theory, 45, 399--431

\bibitem{richardson2001design}
Richardson T J, Shokrollahi M A and Urbanke R L, 2001, Design of Capacity-Approaching Irregular Low-Density Parity-Check Codes, IEEE Transactions on Information Theory, 47, 619--637

\bibitem{baumstark2023accelerating}
Baumstark A, Jibril M A and Sattler K U, 2023, Accelerating Large Table Scan Using Processing-In-Memory Technology, Datenbank-Spektrum, 23, 199--209

\bibitem{hyun2024pathfinding}
Hyun B, Kim T, Lee D and Rhu M, 2024, Pathfinding Future PIM Architectures by Demystifying a Commercial PIM Technology, IEEE International Symposium on High-Performance Computer Architecture, 263--279

\bibitem{friesel2023fullsystem}
Friesel B, Lütke Dreimann M and Spinczyk O, 2023, A Full-System Perspective on UPMEM Performance, Proceedings of the 1st Workshop on Disruptive Memory Systems, 1--7

\bibitem{shi2025dimm}
Shi S, Yang F, Li Z, Li X and Sun N, 2025, Exploring the DIMM PIM Architecture for Accelerating Time Series Analysis, IEEE Computer Architecture Letters, 24, 169--172

\bibitem{yao2023belief}
Yao H, Abu Laban W, H\"ager C, Graell i Amat A and Pfister H D, 2023, Belief Propagation Decoding of Quantum LDPC Codes with Guided Decimation, arXiv preprint, arXiv:2312.10950

\bibitem{gallager1962low}
Gallager R G, 1962, Low-Density Parity-Check codes, IRE Transactions on Information Theory, 8, 21--28

\bibitem{fossorier1999reduced}
Fossorier M P C, Mihaljevi\'c M and Imai H, 1999, Reduced Complexity Iterative Decoding of Low-Density Parity-Check Codes Based on Belief Propagation, IEEE Transactions on Communications, 47, 673--680

\bibitem{chen2002near}
Chen J and Fossorier M P C, 2002, Near optimum universal belief propagation based decoding of Low-Density Parity-Check codes, IEEE Transactions on Communications, 50, 406--414

\bibitem{chen2002density}
Chen J and Fossorier M P C, 2002, Density Evolution for Two Improved BP-Based Decoding Algorithms of LDPC Codes, IEEE Communications Letters, 6, 208--210

\bibitem{chen2005implementation}
Zhao J, Zarkeshvari F H and Banihashemi A H, 2005, On Implementation of Min-Sum Algorithm and Its Modifications for Decoding Low-Density Parity-Check (LDPC) Codes, IEEE Transactions on Communications, 53, 549--554

\bibitem{chen2005improved}
Chen J, Tanner R M, Jones C and Li Y, 2005, Improved min-sum decoding algorithms for irregular LDPC codes, IEEE International Symposium on Information Theory, 449--453

\bibitem{myung2017offset}
Myung S, Park S I, Kim K J, Lee J Y, Kwon S and Kim J, 2017, Offset and Normalized Min-Sum Algorithms for ATSC 3.0 LDPC Decoder, IEEE Transactions on Broadcasting, 63, 734--739

\bibitem{wulf1995hitting}
Wulf W A and McKee S A, 1995, Hitting the memory wall: implications of the obvious, ACM SIGARCH Computer Architecture News, 23, 20--24

\bibitem{devaux2019true}
Devaux F, 2019, The true Processing In Memory accelerator, IEEE Hot Chips 31 Symposium, 1--24

\bibitem{asifuzzaman2023survey}
Asifuzzaman K, Miniskar N R, Young A R, Liu F and Vetter J S, 2023, 
A survey on processing-in-memory techniques: Advances and challenges, 
Memories, Materials, Devices, Circuits and Systems, 4, 100022

\bibitem{barroso2013tail}
Dean J and Barroso L A, 2013, The tail at scale, Communications of the ACM, 56, 74--80

\bibitem{ihler2005loopy}
Ihler A T, Fisher III J W and Willsky A S, 2005, Loopy Belief Propagation: Convergence and Effects of Message Errors, Journal of Machine Learning Research, 6, 905--936

\bibitem{murphy2013loopy}
Murphy K P, Weiss Y and Jordan M I, 1999,
Loopy belief propagation for approximate inference: An empirical study, Proceedings of the Fifteenth Conference on Uncertainty in Artificial Intelligence, 467--475

\bibitem{raveendran2021trapping}
Raveendran N and Vasić B, 2021, Trapping Sets of Quantum LDPC Codes, Quantum, 5, 562

\bibitem{kuo2022exploiting}
Kuo K Y and Lai C Y, 2022, Exploiting degeneracy in belief propagation decoding of quantum codes, npj Quantum Information, 8, 111

\bibitem{haffner2008quantum}
Häffner H, Roos C F and Blatt R, 2008, Quantum computing with trapped ions, Physics Reports, 469, 155--203

\bibitem{bruzewicz2019Trapped}
Bruzewicz C D, Chiaverini J, McConnell R and Sage J M, 2019, Trapped-ion Quantum Computing: Progress and Challenges, Applied Physics Reviews, 6, 021314

\bibitem{battistel2023realtime}
Battistel F, Chamberland C, Johar K, Overwater R W J, Sebastiano F, Skoric L, Ueno Y and Usman M, 2023, Real-Time Decoding for Fault-Tolerant Quantum Computing: Progress, Challenges and Outlook, Nano Futures, 7, 032003

\bibitem{bausch2024learning}
Bausch J, Senior A W, Heras F J H, Edlich T, Davies A, Newman M, Jones C, Satzinger K, Niu M Y, Blackwell S, Holland G, Kafri D, Atalaya J, Gidney C, Hassabis D, Boixo S, Neven H and Kohli P, 2024, Learning high-accuracy error decoding for quantum processors, Nature, 635, 834--840

\bibitem{cross2009comparative}
Cross A W, DiVincenzo D P and Terhal B M, 2009, A comparative code study for quantum fault-tolerance, Quantum Information \& Computation, 9, 541--572

\bibitem{ryananderson2021realization}
Ryan-Anderson C, Bohnet J G, Lee K, Gresh D, Hankin A, Gaebler J P, Francois D, Chernoguzov A, Lucchetti D, Brown N C, Gatterman T M, Halit S K, Gilmore K, Gerber J A, Neyenhuis B, Hayes D and Stutz R P, 2021, Realization of real-time fault-tolerant quantum error correction, Physical Review X, 11, 041058

\bibitem{ferraz2024inmemorybf}
Ferraz O, Falcao G and Silva V, 2024, In-Memory Bit Flipping LDPC Decoding, 2024 32nd European Signal Processing Conference (EUSIPCO), 706--710

\bibitem{ferraz2025inmemorynbldpc}
Ferraz O, Silva V and Falcao G, 2025, In-Memory Non-Binary LDPC Decoding, IEEE Access, 13, 186887--186902

\bibitem{valls2021syndrome}
Valls J, Garcia-Herrero F, Raveendran N and Vasic B, 2021, Syndrome-Based Min-Sum vs OSD-0 Decoders: FPGA Implementation and Analysis for Quantum LDPC Codes, IEEE Access, 9, 138734--138743

\bibitem{bascones2025exploring}
Bascones D, Garcia-Herrero F and Valls J, 2025, Exploring the FPGA and ASIC design space of belief propagation and ordered statistics decoders for quantum error correction codes, EPJ Quantum Technology, 12, 140

\bibitem{ferraz2025gpuqldpc}
Ferraz O, Coutinho B C, Falcao G, Gomes M, Monteiro F A and Silva V, 2025, GPU-Accelerated Syndrome Decoding for Quantum LDPC Codes below the 63 $\mu$s Latency Threshold, Asilomar Conference on Signals, Systems, and Computers, 1878--1882


\end{thebibliography}
\end{document}